# CONFINING FORCES AND STRING FORMATION FROM THE LATTICE

G.S. Bali, K. Schilling, Ch. Schlichter, and A. Wachter
*Fachbereich Physik, Bergische Universität,*
*D 42097 Wuppertal, Germany*
and
*Forschungszentrum Jülich, HLRZ, D 52425 Jülich, Germany*

ABSTRACT

We show the running coupling as derived from the SU(3) $Q\bar{Q}$ potential and discuss preliminary results on spin dependent heavy quark potentials from high statistics lattice simulations of SU(2) gauge theory. The precision suffices to study scaling properties and lattice artifacts (at short distances). We identify the Coulomb like short range interaction as a mixed vector-scalar exchange. We measure flux tube formation between a static $Q\bar{Q}$ pair over physical distances up to 2 fm, with a spatial resolution as small as .05 fm. Consistency with the string picture is found for separation larger than about 1 fm, with a half width of the profile of approximately .7 fm.

## 1. The Lattice Approach

Lattice gauge theory provides a powerful tool to compute non-perturbative properties of strong interactions from the basic QCD Lagrangian, such as the confining force between a quark-antiquark pair. The central potential has been the object of lattice investigations, ever since the seminal work of M. Creutz[1] back in 1979.

Over the last years considerable progress has been accumulated, *both* in computational methods and in the computing power, in particular through the advent of high performance parallel computing devices. In this way, the static central potential, $V_0$, could be computed — from improved Wilson loops — with statistical accuracy of a few per mille in $SU(3)$ gauge theory, on large lattices with high resolution[2]. This level of precision allowed to carry out a direct analysis of the short range interaction in terms of the running coupling[3,4] $\alpha_{q\bar{q}}$ in this theory, as we demonstrate[5] in fig. 1, in units of the string tension $K \approx (440\ \mathrm{MeV})^2$. The full curves correspond to the two-loop running coupling with $\Lambda_R = .66(4)\sqrt{K}$. The dotted curves are expected from a funnel potential.

The computation of the spin dependent potentials presents a much greater challenge to large scale computing, as it requires the summation over a variety of connected plaquette-Wilson loop operators[6] with large time extent. The first pioneering attempts in this direction[7,8] have been made in the mid eighties, and have provided evidence – within the statistical and systematic errors of the period – in favour of the common prejudice that the long range confining force is related to scalar exchange, while the short range Coulombic part is predominantly vector like in nature.

Another, not less demanding problem is the computation of the energy and action density distributions around static sources that could shed light on the mechanism of string formation. Technically speaking, these distributions are extracted from the disconnected correlation between a plaquette and a Wilson loop whose time extent, $T$, should be asymptotically large such as to exhibit ground state dominance.In order to study string formation, one needs to realize spatial separations beyond one fermi; in this respect, previous lattice studies were hampered either by too coarse lattice resolution or by too small source separations[9,10].

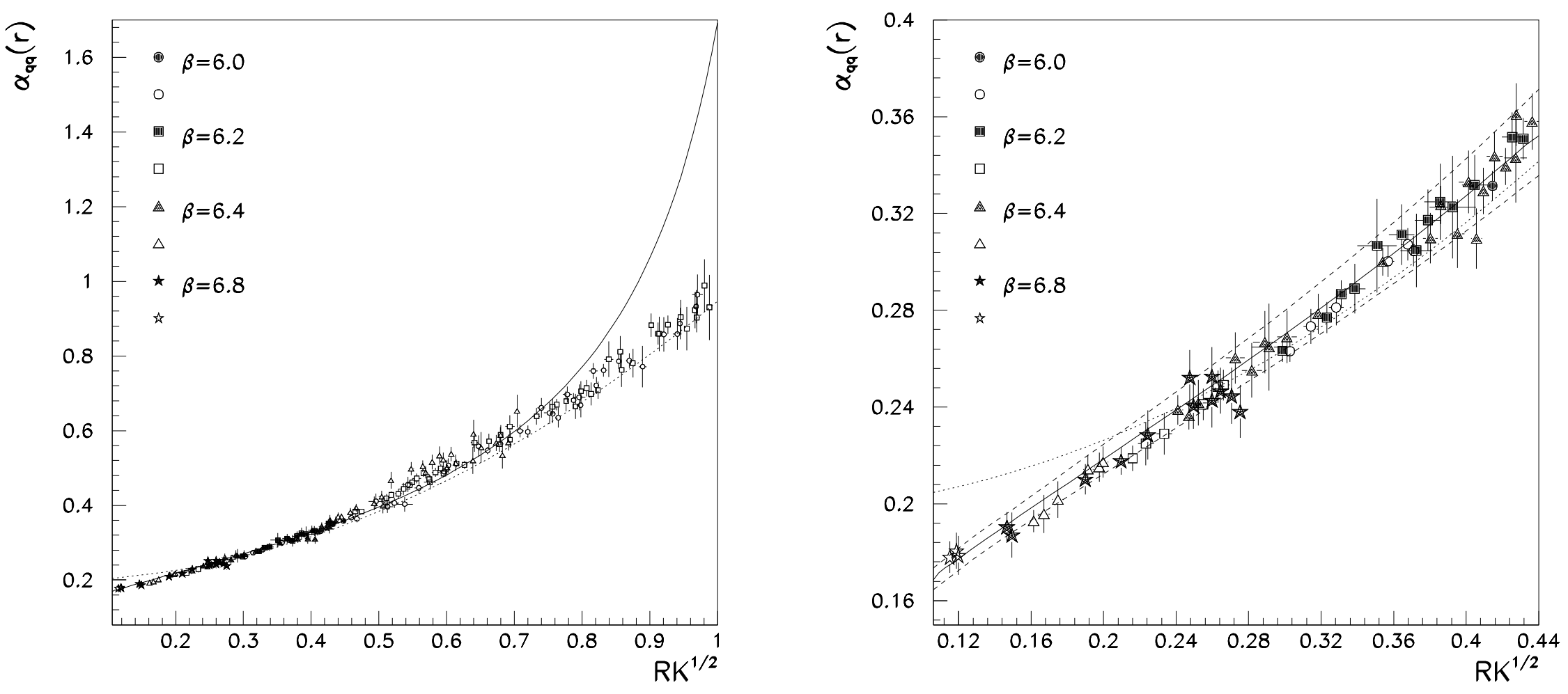

Fig. 1. (a) The running coupling from the interquark force. (b) Same as fig. 1a for $1/r <$ 1 GeV.

In order to produce decent signal to noise ratios from these lattice measurements, it is crucial to ensure early $T$-asymptotics in the observed correlators. This can be achieved by a suitable "smearing" of the spatial parts in the Wilson loops[11,2,4]. In this contribution we shall present an outline of the physics results from our recent high statistics[12,13] investigations, where we applied such smearing techniques. For the sake of statistics, we base our work on $SU(2)$ gauge theory as it is easier to handle on the computer and has a vacuum structure similar to the $SU(3)$ case.

## 2. Spin Dependent Potentials

A $1/m$ expansion of the heavy quark potential in the equal mass case yields the Breit-Fermi type expression (assuming spin independence of the leading order potential, $V_0$),

$$V_{SD}(r) = V_0(r) + \frac{1}{m^2}\left(V_{LS}(r) + V_{SS}(r) + \text{spin independent corrections}\right) + \cdots \quad (1)$$

with

$$V_{LS}(r) = \frac{\vec{L}_1\vec{s}_1 - \vec{L}_2\vec{s}_2}{r}\left(\frac{V_0'(r)}{2} + V_1'(r)\right) + \frac{\vec{L}_1\vec{s}_2 - \vec{L}_2\vec{s}_1}{r}V_2'(r) \quad , \quad (2)$$

$$V_{SS}(r) = \left((\hat{r}\vec{s}_1)(\hat{r}\vec{s}_2) - \frac{\vec{s}_1\vec{s}_2}{3}\right)V_3(r) + \frac{\vec{s}_1\vec{s}_2}{3}V_4(r) \quad , \quad (3)$$

where $\vec{L}_i = \vec{r} \times \vec{p}_i$. For Dirac fermions explicit expectation values have been associated to the spin-orbit and spin-spin "potentials" $V_1', V_2'$ and $V_3, V_4$, which can be computed in form of "eared" Wilson loops on the lattice[6,14]. These potentials are related to scalar ($S$), vector ($V$) and pseudo-scalar ($P$) exchange contributions in the following way:

$$V_0(r) = S(r) + V(r) \tag{4}$$

$$V_3(r) = \frac{V'(r) - P'(r)}{r} - (V''(r) - P''(r)) \tag{5}$$

$$V_4(r) = 2\nabla^2 V(r) + \nabla^2 P(r) \quad . \tag{6}$$

In addition, relativistic invariance leads to the Gromes relation[15]

$$V_0' = V_2' - V_1' \quad . \tag{7}$$

The central potential, $V_0$, has a very clear linear confining part (see fig. 2a) which must be of scalar exchange type as a vector exchange can have at most a logarithmic asymptotic $r$ dependence[16]. The lattice potentials $V_1^L$ to $V_4^L$ undergo multiplicative renormalizations in respect to their continuum counterparts, while $V_0$ does not. We have performed a non-perturbative renormalization of these quantities in the manner suggested by Michael *et al.*[17]. In fig. 2b we compare the Gromes combination $V_2' - V_1'$ (in units of the string tension, $K$) to the force, derived from the fit to the central potential as given in fig. 2a. Two data sets for the $\beta$-values 2.74 and 2.96 (lattice resolutions .04 fm and .02 fm) are combined: we find good scaling behaviour and agreement with the Gromes relation outside of the region of lattice artifacts. This demonstrates the success of Michael's renormalization prescription.

We confirm the second spin-orbit potential (see fig. 3b) to be definitely of short range nature, leaving little room for a scalar contribution to this potential. The first spin-orbit potential, $V_1'$, is smaller by a factor five and more noisy (see fig. 3a). Our lattice resolution enables us to establish an attractive short range contribution that can be well fitted to a Coulomb ($1/R^2$) form, in addition to the constant long-range term, that is in agreement with the string tension, $K$.

In principle, the potentials $V_0$, $V_3$, and $V_4$ allow for a determination of $S$, $V$, and $P$. At this stage, we will assume $P$ to vanish and $V_1$ to be pure scalar (which induces the equalities $V_2 = V$ and $V_1 = -S$). This leads to a prediction for $V_3$, according to eq. (5), which can be checked in fig. 4a: we find reasonable agreement between the data sets and the predicted curve, $V_2'/R - V_2''$, the deviations being qualitatively understood from tree level lattice perturbation theory.

The remaining spin-spin potential, $V_4$, exhibits oscillatory behaviour as a lattice artifact (see fig. 4b) and can largely be understood as a $\delta$-contribution, according to

$$V_2 = -\frac{c}{r} \Rightarrow V_4 = 2\nabla^2 V_2 = 8\pi c \delta_L^3(\vec{R}) \quad , \tag{8}$$

where $\delta_L$ is the appropriate lattice $\delta$-function for the chosen (symmetric) ear combination of our lattice operator. This expectation is indicated by open squares in the

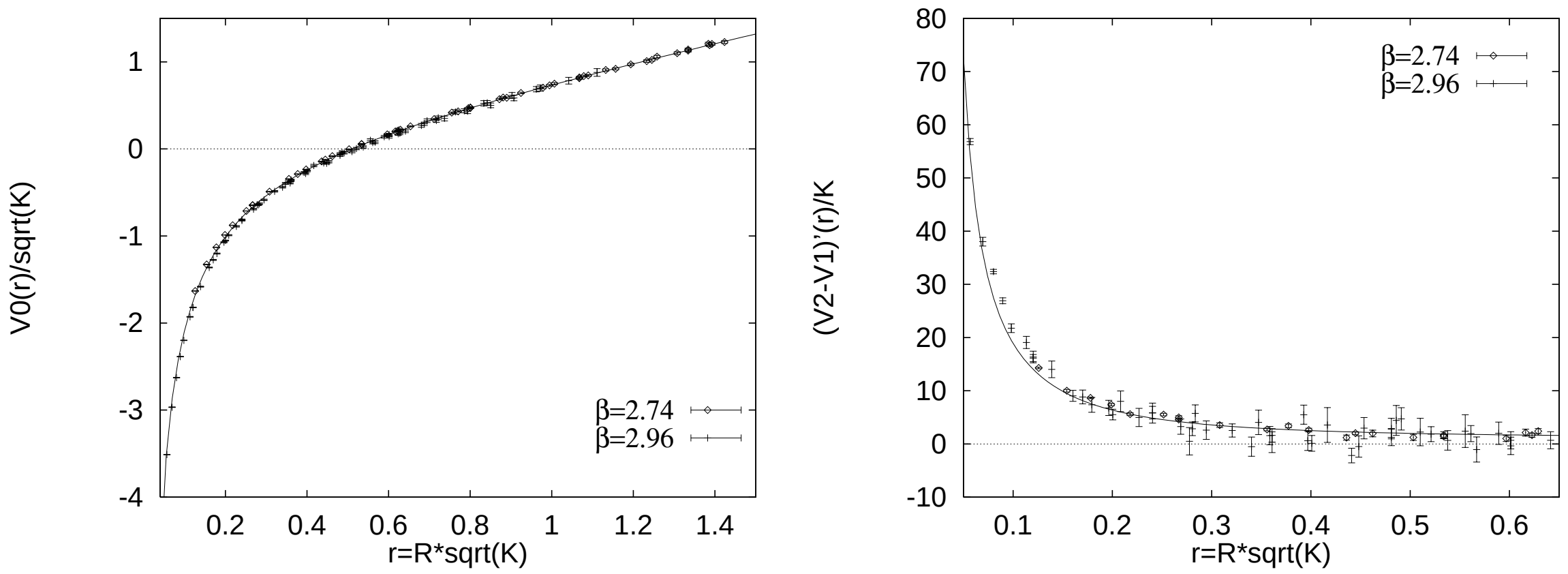

Fig. 2. (a) Central potential with a funnel-type fit curve. (b) Comparison of $V_2' - V_1'$ (data) with the (fitted) central force (curve) (Gromes relation).

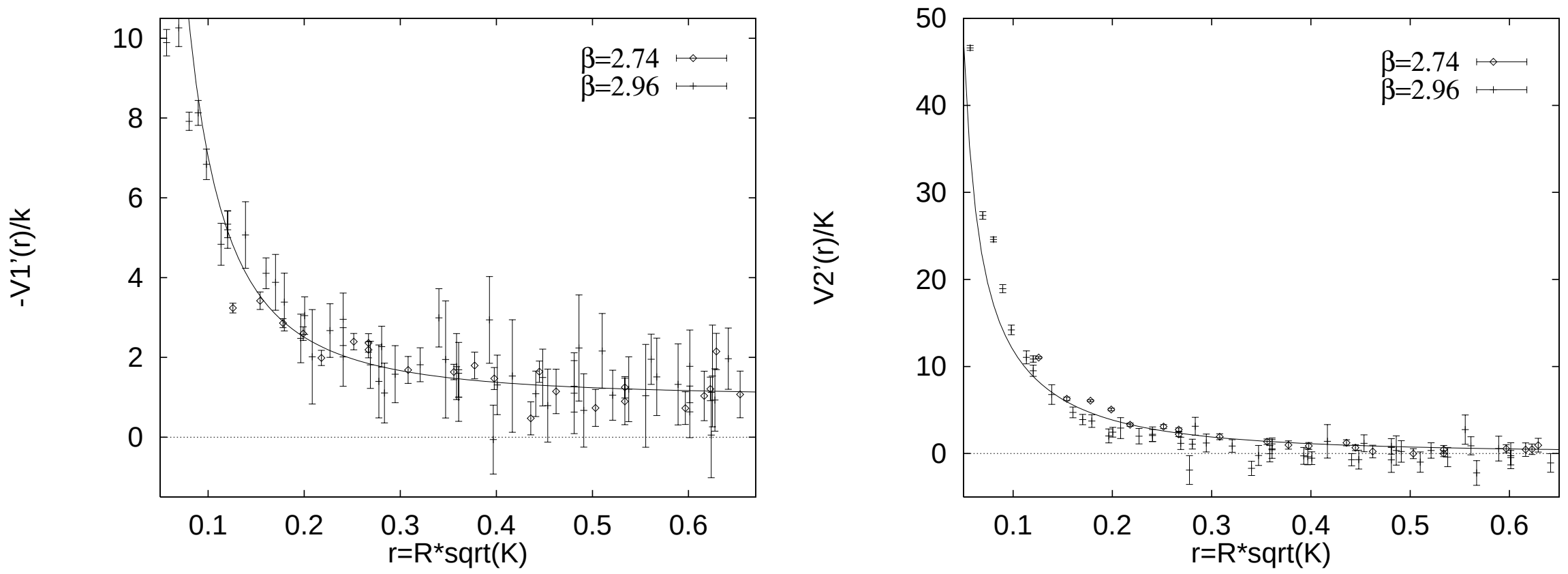

Fig. 3. (a) $V_1'$, together with a fit curve of the form $a/r^2 + K$. (b) The spin-orbit potential $V_2'$. The curve corresponds to the central force plus the (fitted) $V_1'$.

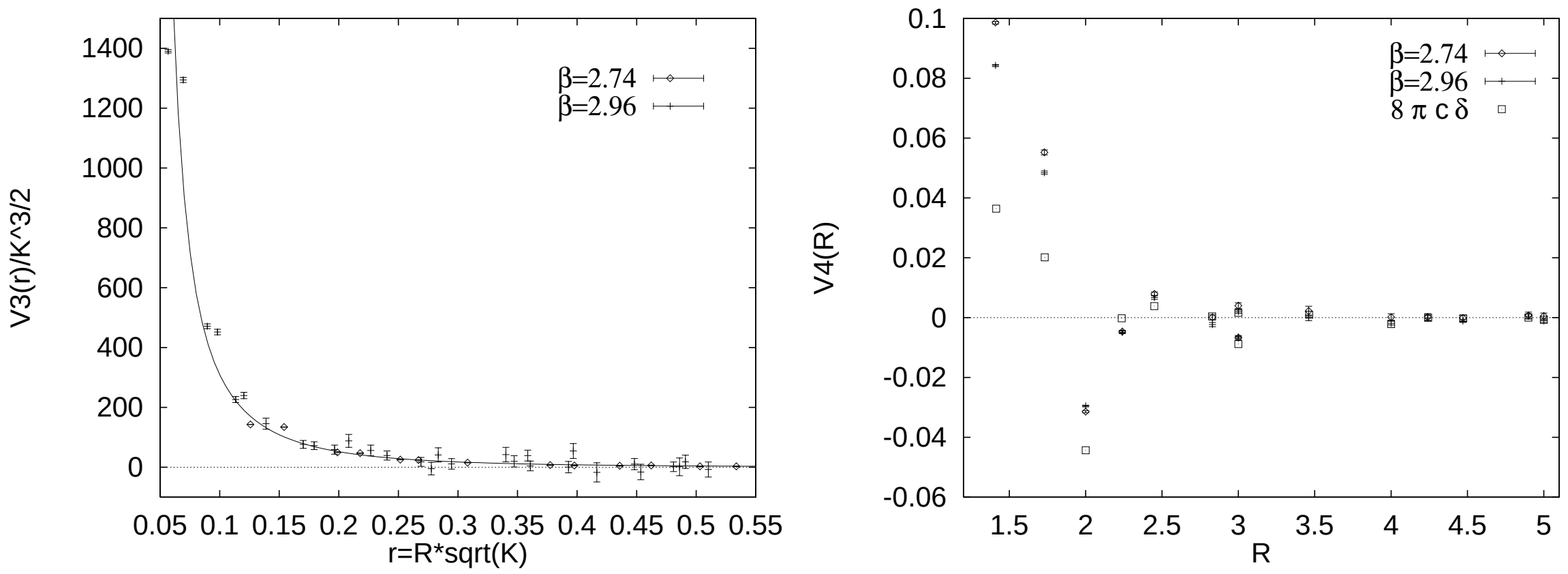

Fig. 4. (a) Comparison of $V_3$ with $V_2'/r - V_2''$ (curve). (b) $V_4$ for the two $\beta$-values in lattice units (points with error bars).The squares indicate the appropriate lattice $\delta$ function with the coefficient, $c$, taken from the $V_2'$ analysis.

figure. Notice, that we have plotted $V_4$ in lattice units on both axes! At $R \leq 2$ we observe an additional $1/R^4$ contribution.

In conclusion, we find a consistent picture when identifying $-V_1$ ($V_2$) with the scalar (vector) exchange contribution, other exchange types being negligible to order $1/m^2$. Apart from the linear large distance part, we observe a short distance Coulomb like scalar contribution from the first spin-orbit potential. It appears that the Coulomb part of the central potential splits up into a vector/scalar ratio in between 3/1 and 4/1.

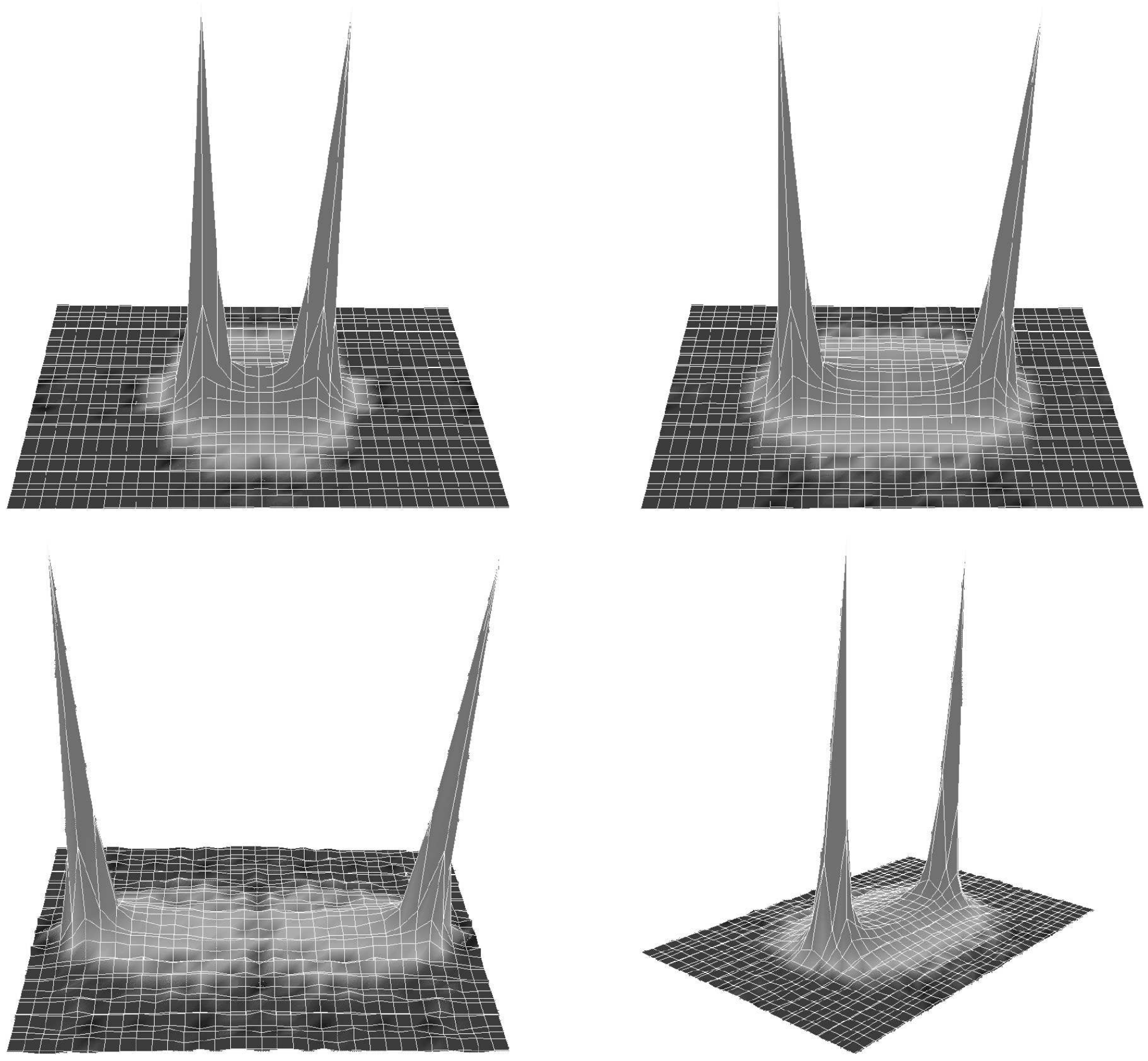

Fig. 5. The action density distribution for various quark separations, measured at $\beta = 2.5$.

## 3. String Formation

We are able to map out the action density distributions around static SU(2) sources with separations, $r$, ranging up to 2 fm[12]. Figs. 5a to 5c illustrate the formation of an elongated string by varying the source separation from .6 over 1.1 to 2 fm, the largest distance corresponding to 24 lattice spacings at $\beta = 2.5$. Fig. 5d contains the information of fig. 5b, but observed from a different angle of view, in order to exhibit the width of the string. Notice, that the plots represent the data without any interpolation. The brightness has been computed from the relative statistical errors.*

We observe a linear widening of the width of the flux tube with the quark separation, $r$, up to $r \approx .7$ fm. After saturation, from $r \approx 1$ fm onwards, the width remains rather constant. The transverse profile is in agreement with both, a dipole type and a Gaussian shape, within statistical accuracy. We find a plateau value, $\rho \approx .7$ fm, for the half width. The corresponding mean squared radius, $\delta$, is found to be within the range $.5 \text{ fm} < \delta < .8$ fm.

## 4. Acknowledgements

This project was supported by DFG (grant Schi 257/3-1) and by EU (grant SC1*-CT91-0642). We thank HLRZ for the computer time on their CM5 system.

*A data base of colour images of the flux tubes and related quantities can be accessed via anonymous ftp from wpts0.physik.uni-wuppertal.de. The (compressed) .rgb and .ps files can be found in the directory pub/colorflux.